\documentclass[preprint,showpacs]{revtex4-1}
\usepackage[caption=false]{subfig}
\usepackage{graphicx,psfrag}
\usepackage{placeins}
\usepackage{amsfonts}
\usepackage{fancyhdr}
\usepackage{setspace}
\usepackage{mathrsfs}
\usepackage{xcolor}

\usepackage{amsmath,amssymb,enumerate}
\usepackage{ulem}
\begin{document}

\title[Characterizing network topology using first-passage analysis]{Characterizing network topology using first-passage analysis}

\author{M. S. Chaves}
\email{Corresponding author: marcelochaves.chaves@gmail.com}
\affiliation{Programa de P\'{o}s-Gradua\c{c}\~{a}o em Modelagem Matem\'{a}tica e Computacional, Centro Federal de Educa\c{c}\~{a}o Tecnol\'{o}gica de Minas Gerais - CEFET-MG, Brazil}
\author{T. G. Mattos}
\affiliation{Departamento de F\'{\i}sica, Centro Federal de Educa\c{c}\~{a}o Tecnol\'{o}gica de Minas Gerais - CEFET-MG, Av. Amazonas 7675, 30.510-000, Belo Horizonte-MG, Brazil}
\author{A. P. F. Atman}
\affiliation{Departamento de F\'isica - Centro Federal de Educa\c c\~ao Tecnol\'ogica de Minas Gerais, CEFET-MG and Instituto Nacional de Ci\^encia e Tecnologia - Sistemas Complexos.}

%\affiliation{Departamento de F\'{\i}sica e Matem\'{a}tica and National Institute of Science and Technology for Complex Systems, Centro Federal de Educa\c{c}\~{a}o Tecnol\'{o}gica de Minas Gerais - CEFET-MG, Av. Amazonas 7675, 30.510-000, Belo Horizonte-MG, Brazil}

\begin{abstract}
Understanding the topological characteristics of complex networks and how they affect navigability is one of the most important goals in science today, as it plays a central role in various economic, biological, ecological and social systems. Here, we apply First Passage analysis tools to investigate the properties and characteristics of random walkers in networks with different topology. Starting with the simplest two-dimensional square lattice, we modify its topology incrementally by randomly reconnecting links between sites. We characterize these networks by First Passage Time from a significant number of random walkers without interaction, varying the departure and arrival locations. We also apply the concept of First Passage Simultaneity, which measures the likelihood of two walkers reaching their destination together. These measures, together with the site occupancy statistics during the processes, allowed to differentiate the studied networks, especially the random networks from the scale-free networks, by their navigability. We also show that small world features can also be highlighted with the proposed technique.
\end{abstract}

\date{\today}
\pacs{02.50.-r; 03.65.Nk; 42.25.Dd; 73.23.-b}

\maketitle

\section{Introduction}
\label{sec:intro}

Modern society is undergoing a major change since it has become a massively connected community, forming complex networks with distinctive features, such as small world (SW) phenomena, \cite{Milgram1967,Mitchell2009,Newman2018}. Thus, identifying, characterizing the topology, and understanding the dynamics and navigability properties of these networks have become essential objectives of the research community ~\cite{Goh_2001,Hwang_2013,Watts_1998,Hwang_2014,Erdos1959,Erdos1960,McGraw_2008,Barabasi1999,BARABASI2002,Milgram1967,Gilbert_1959}. Currently, a significant effort has been made to quantify and improve the efficiency of these networks \cite{Goncalves2019,Carpi2019,Oliveira_2019}.
	
Beyond the complex network phenomena, in last years an increasing interest has been addressed to First Passage Phenomena (FPP). FPP underlies a great variety of processes in nature and in human activities and has enormous potential for applications~\cite{FPP2014}. In Biology, for example, statistical analysis of first-passage times for stem-cell ageing models allows for a better understanding of cellular mutation and disease spreading~\cite{Chou2014}. Recently, contingent convertible bond pricing methods have been derived from analysis based on two-dimensional stochastic processes~\cite{Choe2019}. The proposed dynamic capital-ratio model consider that the stock price follows a geometric Brownian motion, and the first-passage time is defined as the stopping time, \textit{i.e.}, when the capital-ratio attains a certain value. Another interesting application of first-passage analysis is the integrate-and-fire model~\cite{Burkitt_2006}, in which a neuron fires only when a floating voltage reaches a specific level for the first time.

Within the wide range of applications, there is one feature that is always present: all systems involved are structured in discrete elements interacting across a network. In such context, a central question to be addressed is how to quantitatively characterize the topology of a network. To that end, one is often concerned about statistical properties of the node distribution, as well as the nature of the connections between them. Additionally, one might be also interested in characterizing the navigability of the network, as well as the dynamics of link formation and interaction between nodes. A robust tool for this purpose is the analysis of Brownian motion along the network~\cite{Tejedor_2009, Tejedor_2010, Rosvall_2008, Montroll1964, Montroll_1965, Condamin2005, Benichou2010}.

Advances in complex networks theory  \cite{Erdos1959,Barabasi1999} have increased the interest in applying the ideas of FPP analysis on networks, such as the exact expression for the mean first-passage time (MFPT) between two nodes on the network \cite{Noh_2004}. In Ref.~\cite{Tejedor_2009}, for example, the authors have shown that there is a lower bound for the MFPT of a random walk to a target site averaged over its starting position.
	
In this paper we present a new technique to characterize quantitatively the topology of complex networks. It is based on the analysis of first-passage properties and of sample-to-sample fluctuations of random walks on such networks. The rest of this paper is organized as follows. In Sec.~\ref{sec:methodology}, we describe the method employed to generate the networks, as well as the tools used to perform statistical analysis of first-passage related quantities. In Sec.~\ref{sec:results}, we discuss the results obtained for a few types of networks with different boundary conditions. Finally, in Sec.~\ref{sec:concl}, we draw some conclusions.

\section{Methodology}\label{sec:methodology}

In this section we present the methodology to build the different types of networks and, in the next subsection, the FPP tools employed to analyze the data.

	\subsection{Generating the networks}\label{subsec_networks}
		
We consider a few types of networks and analyze the effects of rewiring between their sites. Real networks are complex dynamic systems that are constantly evolving. Adding or removing elements can cause significant changes to their structures. In addition, new relationships between members may emerge or be extinguished. Depending on how these relationships are established, significant changes between the communication of their elements may occur. To better understand these topological network changes, we consider different rebinding methods to produce different types of complex networks, as described below.

We start with a square lattice ($N=10^4$ sites), with von Neumann neighborhood~\cite{Margolus1987}, and proceed by performing rewiring operations in order to obtain each of the following types of network: conservative random network, non-conservative random network and scale-free network. In each case, we apply the following rules for rebinding links: (i) each site must keep at least one link and (ii) rebinds cannot reinstate the original links.

	\begin{figure}[!htbp] %h or !htbp
		\subfloat[Square network.]{\includegraphics[width=0.5\linewidth]{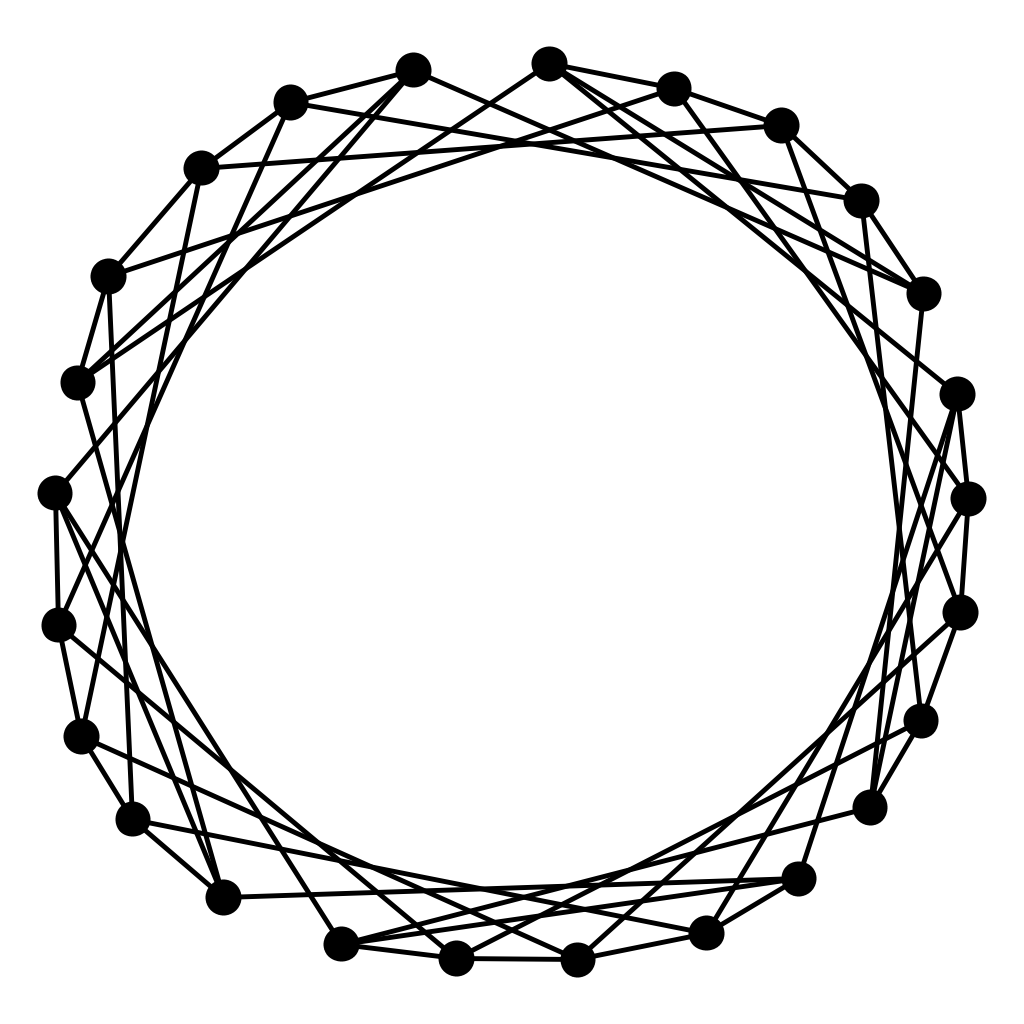}}
		\subfloat[Conservative random network.]{\includegraphics[width=0.5\linewidth]{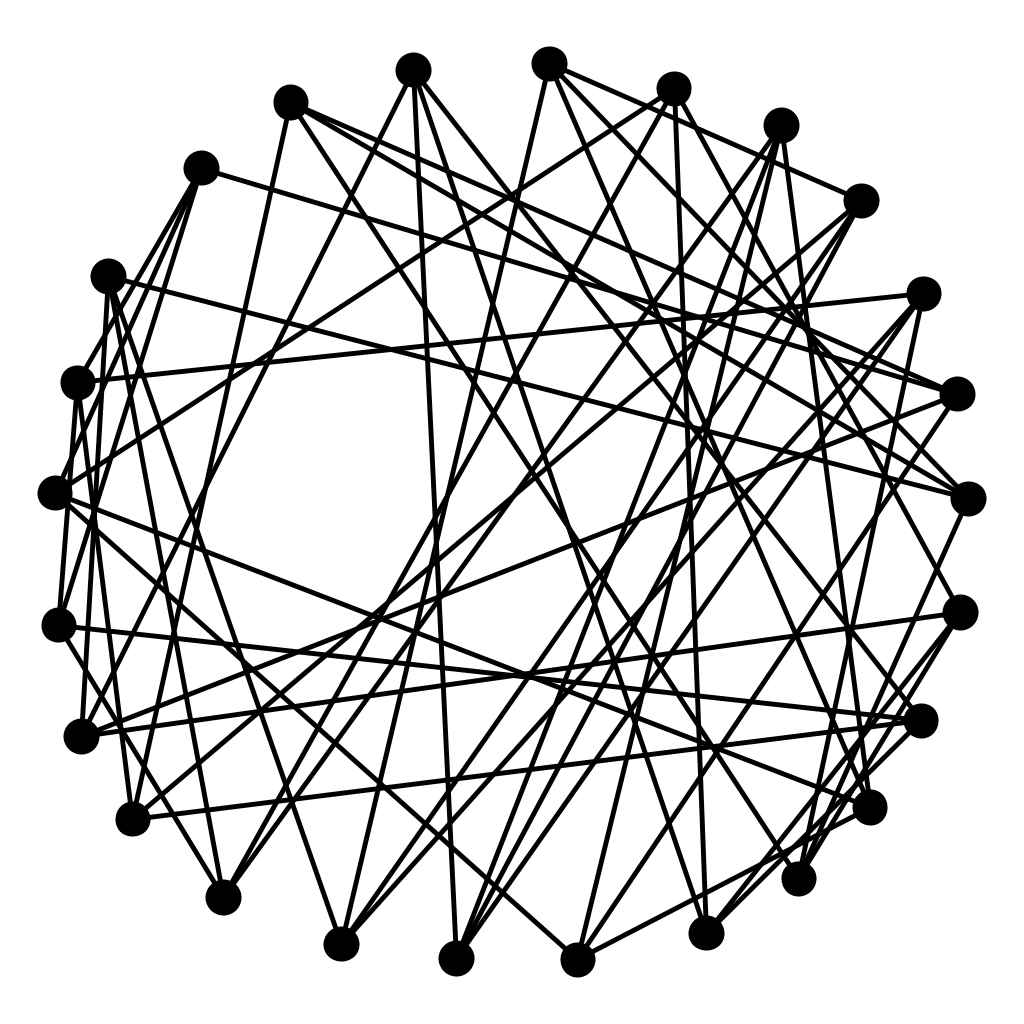}} \\
		\subfloat[Non-conservative random network.]{\includegraphics[width=0.5\linewidth]{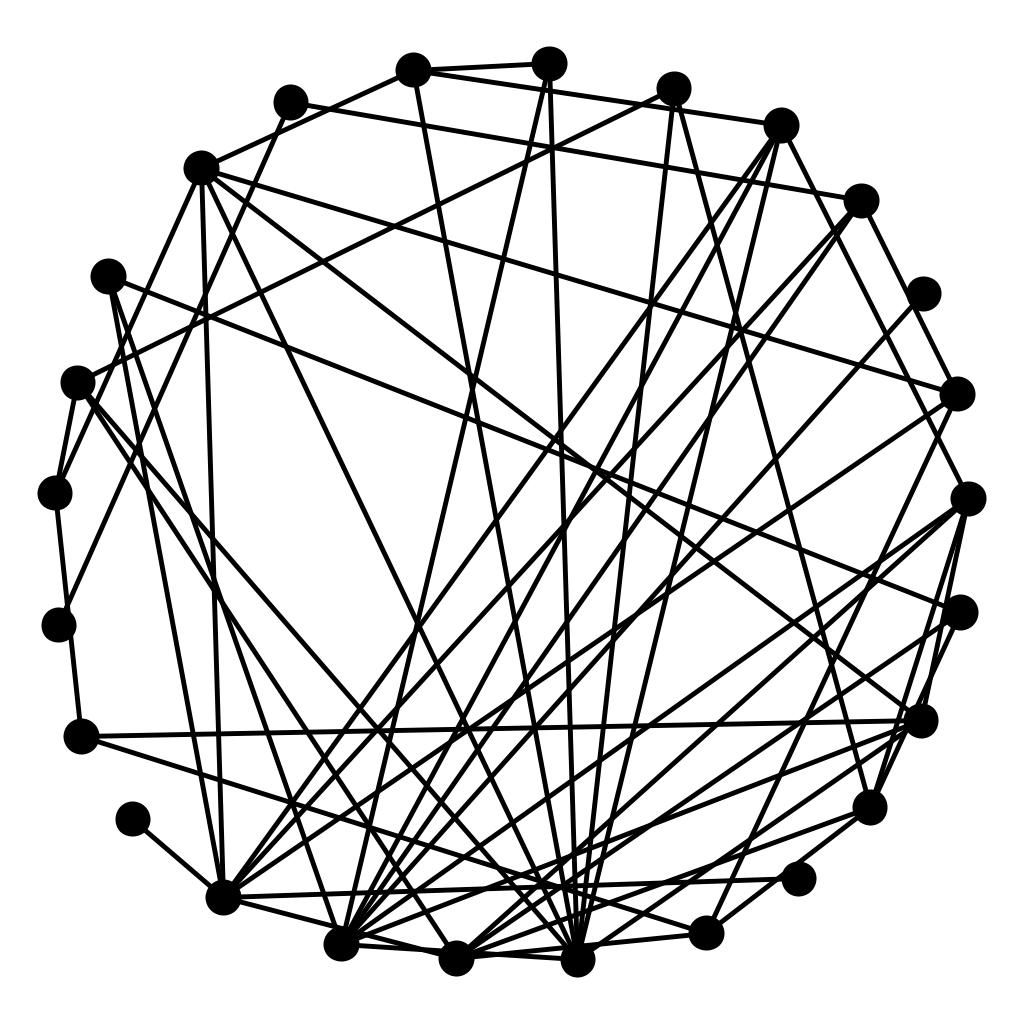}}
		\caption{(a) Square network ($L$$\times$$L$, $L$=5) with periodic boundary condition. Each site has 4 links and the border sites are connected to the symmetric sites of the opposite border. (b) Conservative random network. Starting from a square network two unconnected sites are randomly chosen and a new connection is established between them. A site adjacent to each of these sites is chosen randomly to disconnect from these sites and establish a new connection. These steps are repeated until all the links to be rewiring and at the end, each site keeps 4 links. (c) Non-conservative random network. Starting from a square network two unconnected sites are randomly chosen and a new connection is established between them. A single site next to one of these sites is chosen to lose a link. Theses steps are typically repeat to approximately $90\%$ of original links.}
		\label{lattices}
	\end{figure}
    
  To get the conservative random network, we reconnect the square network links by randomly choosing two unconnected sites and connecting them. To do this, a random link from each site is chosen to be rewired. If these sites are not already connected, we will connect them. Otherwise, we choose two other links to reconnect. Thus, the degree of each location remains at 4 - Fig. \ref{lattices} (b). This operation is repeated until all links in a square network are reconnected, as shown in Fig. \ref{lattices} (b).
    
  For non-conservative random network topology, we randomly choose 2 sites to connect whether they have no links in common, and disconnect one link from each site, also randomly. We continue this operation until the desired number of reconnections is performed. It is noteworthy that we are careful not to allow any site to be disconnected from the network, ensuring that the final network is connected - Fig. \ref{lattices} (c). Typically, the algorithm executes up to approximately $ 90 \% $ of the original square network links has been  removed - Fig. \ref{lattices} (c).

	\begin{figure}[!h] %h or !htbp
		\subfloat[0 Rewiring.]{\includegraphics[width=0.55\linewidth]{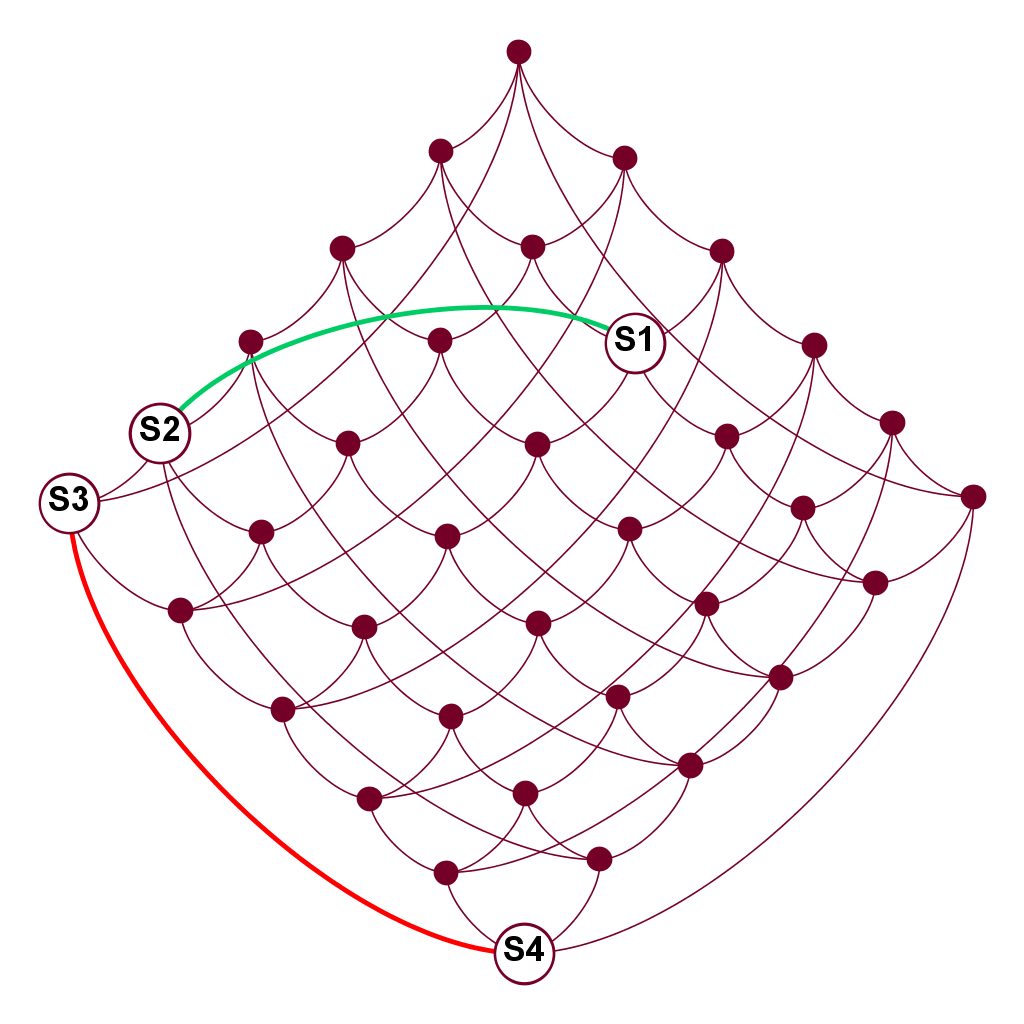}}
		\subfloat[20 Rewiring.]{\includegraphics[width=0.55\linewidth]{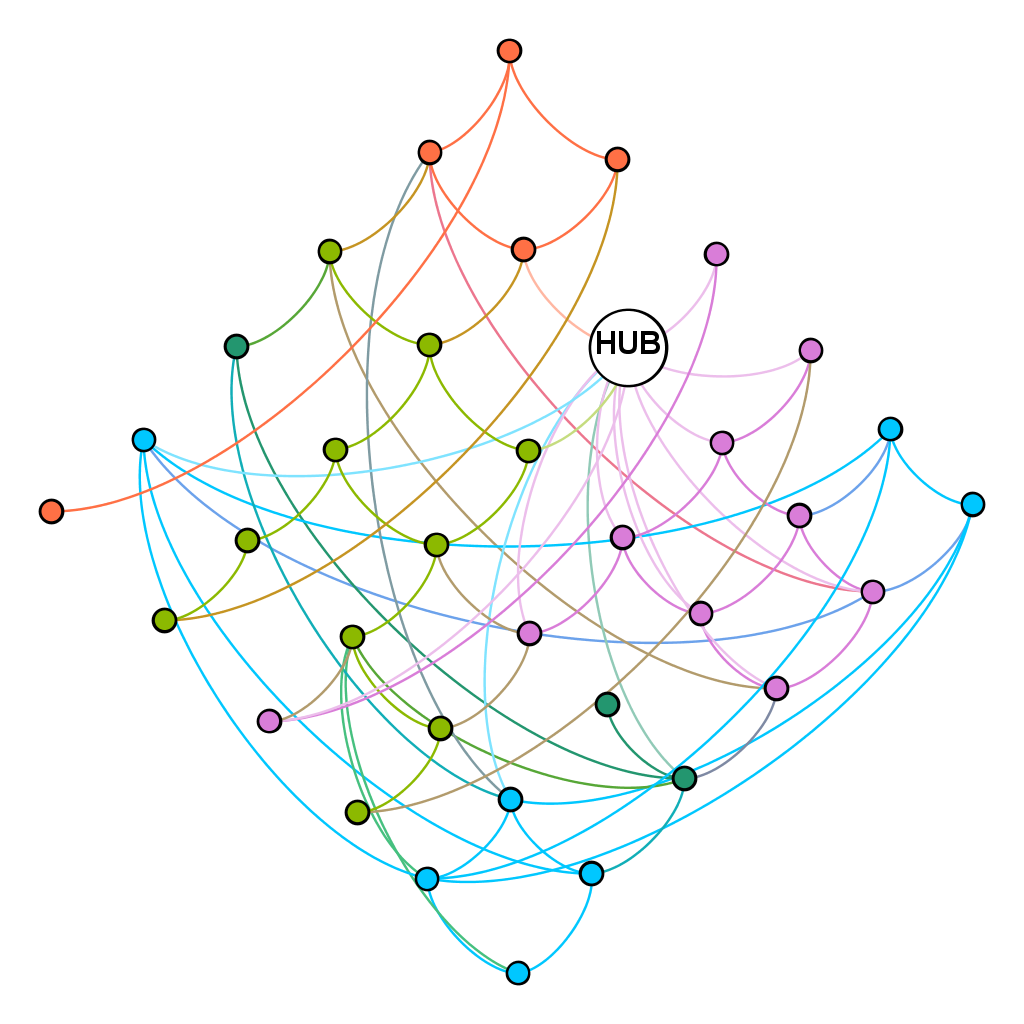}} \\
		\subfloat[66 Rewiring.]{\includegraphics[width=0.7\linewidth]{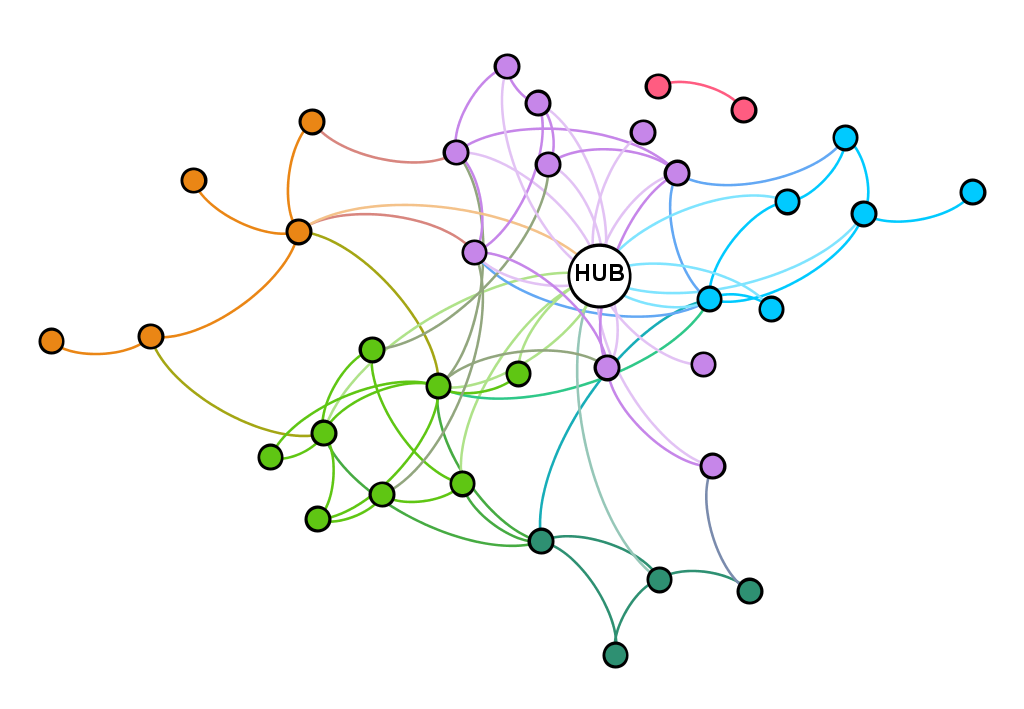}}

		\caption{(Color online) Rewiring a square network using the preferential attachment. (a) Square network ($L$$\times$$L$, $L$=5) with periodic boundary condition. Each site has 4 links and the border sites are connected to the symmetric sites of the opposite border. In the first step $S1$ and $S2$ are connect, and $S3$ and $S4$ will be disconnect. (b)  After 20 rewiring, the site $S1$ earned more links than the others. (c) With approximately $90\%$ of links rewired, the site $S1$ is the hub of network.}
		\label{square_scale_free}
	\end{figure}
	
	\begin{figure*}[htb] %h or !htbp
		%\centering
		\begin{tabular}{cc}
			\subfloat[1000 rewiring.]{\includegraphics[width=0.495\linewidth]{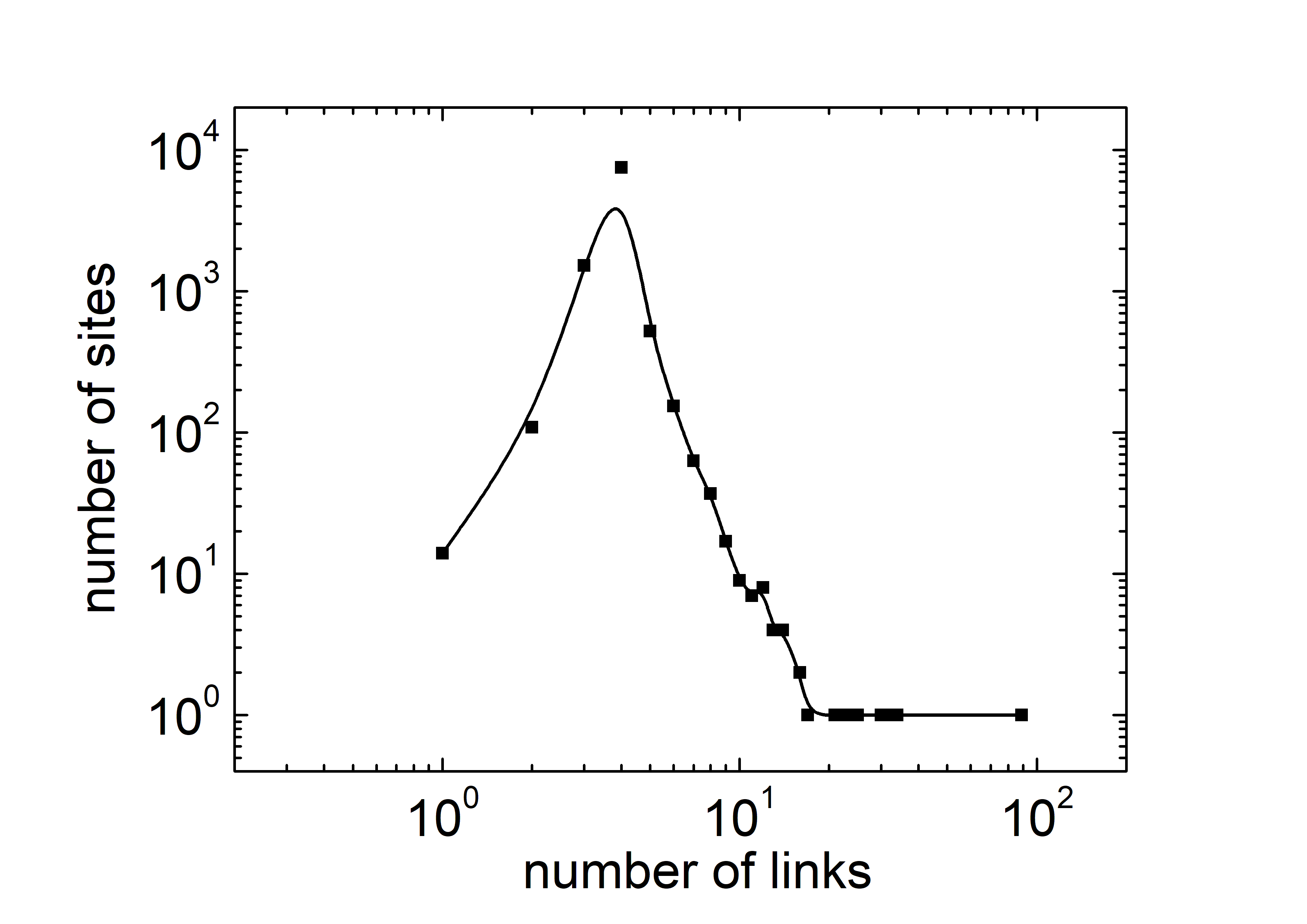}}&
			\subfloat[18577 rewiring.]{\includegraphics[width=0.495\linewidth]{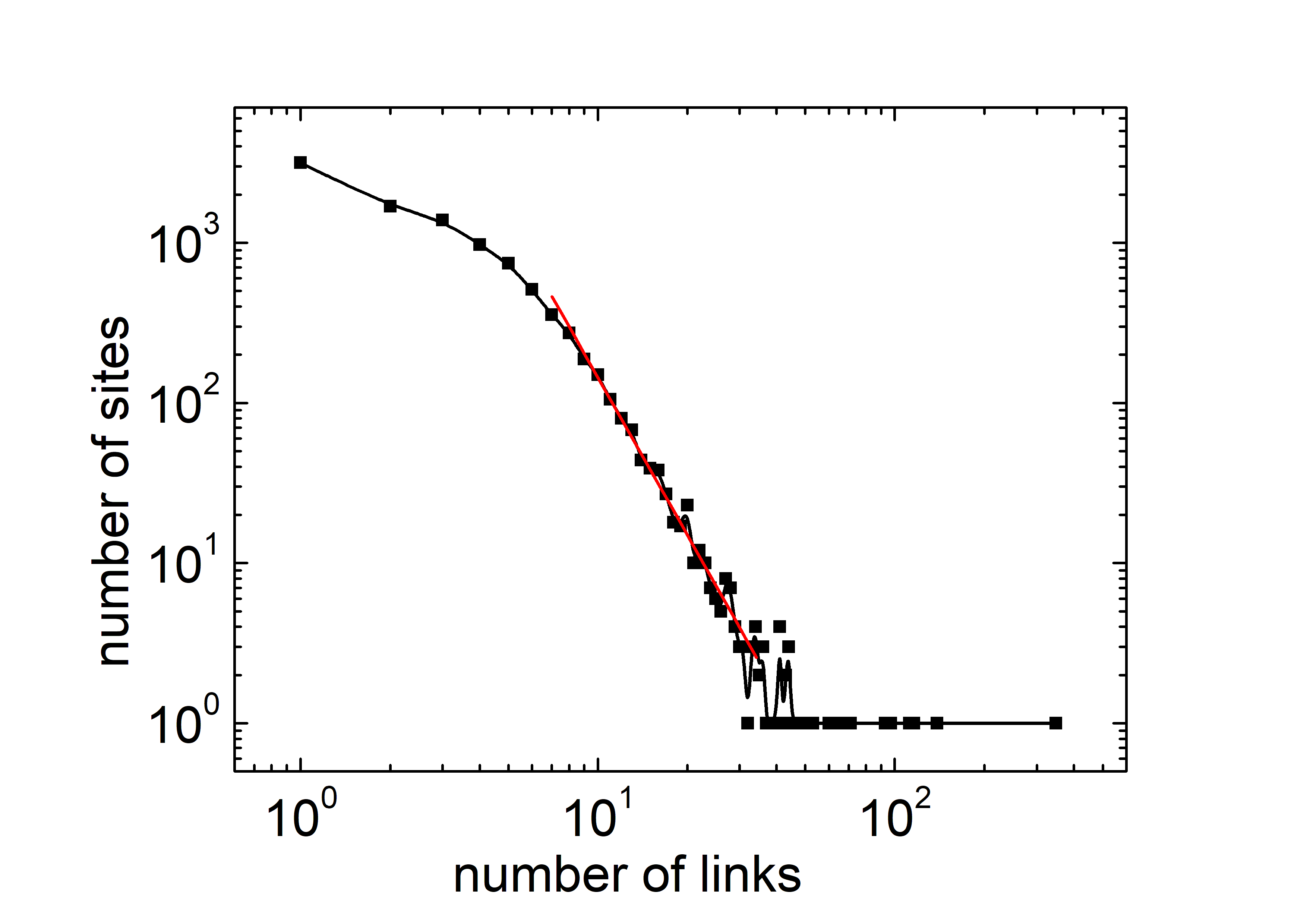}}
		\end{tabular}	
		\caption{Distribution of links during the topological change from square network to scale-free network topology. (a)1000 links and (b) 18577 links were rewired.}
		\label{links}
	\end{figure*}
	
	For obtain scale-free network, we reconnect the square network links using the preferential attachment \cite{Barabasi1999}, so we assume that the probability $P$ of a new connection to site $S_{i}$ depends on its connectivity $k_{i}$, so that $P(k_{i})=k_{i}/P(k_{i})=k_{i}/\sum_{j}k_{j}$. We randomly choose two sites to connect, if they don't have links that link them. Thus, these two sites are more likely to receive new links from other sites in the network. Then all sites are traversed and, according to their connectivity $ k_ {i} $ can get a new connection from a randomly chosen site. For each new link one of original link is disconnected. We continue this operation until the desired number of reconnections is performed. Typically, the algorithm executes up to approximately $ 90 \% $ of the original square network links has been  removed, Fig. \ref{square_scale_free} (c). As the network topology changes, the format of the link distribution changes. Initially, there is a peak centered at 4 original neighbors, but performing rewiring of links the distribution becomes scale-free; see Fig. \ref{links}.
	
	We implemented the SW random network proposed by Watts-Strogatz \cite{Watts_1998}, starting with a ring of $10^4$ sites and 8 non-directed links by site. Clockwise, each site has an original link reconnected to another network site.These procedure are repeat until the desired number of rewires.

	\subsection{First-passage analysis}
	
	Once we have generated a network as described in subsec.~\ref{subsec_networks}, we perform random walks (RWs) on it. The walk consists of random jumps between nearest-neighbor sites. We let the walker start on a determined site of the lattice and record its first-passage time $\tau$ to another given site of the network.
	
	The first-passage time $\tau$ is a random variable defined as the number of steps executed by a random walker when it reaches a given target for the first time. The distribution $\Psi (\tau)$ can be obtained exactly only for very few special situations~\cite{Mejia-Monasterio2011} and one must often resort to computer simulations.
	
	Furthermore, $\Psi (\tau)$ depends on several topological features, such as geometry, dimensionality, and boundary conditions~\cite{Havlin_2002,Redner2007,Reis2014}. For RWs in unbounded domains $\Psi (\tau)$ is typically broad, i.e., it decays as a power law. In such case it does not possess all moments, even the first moment (i.e., the mean first-passage time $\left< \tau\right>$). 
	
	This behavior is also observed in stochastic dynamics in which scale-free waiting times or topologies are considered~\cite{Metzler_2000,Scher_2002}. In such cases, typical statistical quantities such as the mean and the variance do not yield useful information regarding sample to sample fluctuations, which are important if one is interested in determining the efficiency of search processes, for example~\cite{Mejia-Monasterio2011}.
		
	On the other hand, for RWs in bounded domains $\Psi (\tau)$ is typically a narrow distribution, possessing moments of arbitrary order. In such cases, one is often interested in standard statistical quantities such as, e.g., averages, standard deviations, skewness, and kurtosis.
	
	In order to address this issue, we refer to the concept of simultaneity, which corresponds to the event in which two independent brownian particles arrive at an absorbing boundary at the same time, given that they have departed from the same site. We define the uniformity index

	\begin{equation}\label{omega_def}
	\omega = \dfrac{\tau_i}{\tau_i+\tau_j}\,.
	\end{equation}
	
	$\omega$ is a random variable in the interval $[0,1]$ which measures the likelihood that two independent random walkers departing from the same site reach a target at the same time. If the distribution $P(\omega)$ is bell-shaped, with a maximum at $\omega = 1/2$, then it is likely that two independent walkers will reach the target at the same time and the dynamics is uniform.
	
	On the other hand, if $P(\omega)$ is M-shaped, with a minimum at $\omega = 1/2$, the system is strongly non-uniform and sample-to-sample fluctuations play a key role in the dynamics~\cite{Mejia-Monasterio2011,Mattos_2012,Mattos_2014}.
	
	Since the shape of $P(\omega)$ depends on topological characteristics of the domain, it can be used to identify the topology of a network.
	
	\section{Results} \label{sec:results}
\begin{figure}[!htp]
		
		\subfloat[]{\includegraphics[width=0.7\linewidth]{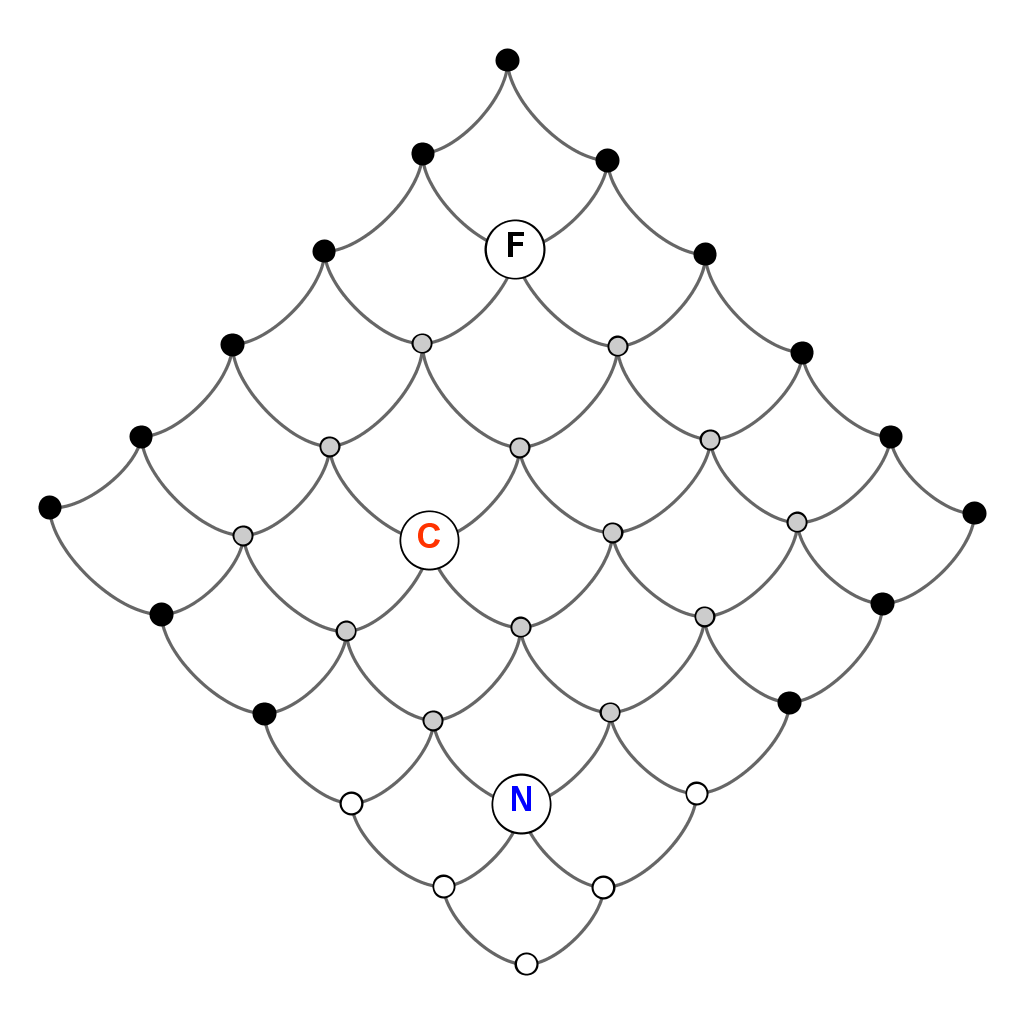}}	\\
		\subfloat[]{\includegraphics[width=1.0\linewidth]{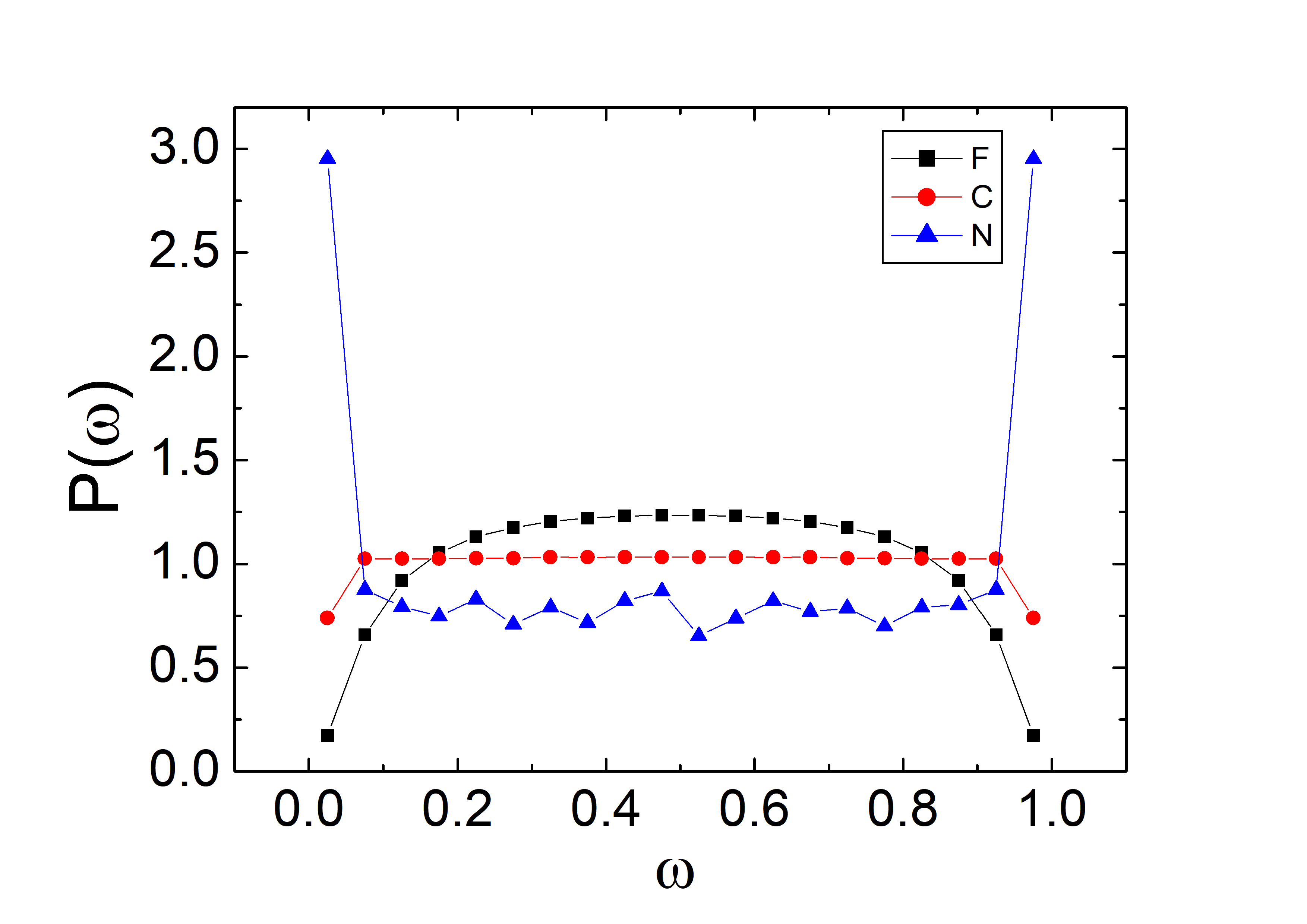}} 		
		\caption{(Color online) (a) Square lattice with mixed BCs: black dots correspond to reflecting boundaries and white dots correspond to absorbing sites. The nodes highlighted correspond to: F (far from the absorbing boundary), C (central region of the lattice), and N (near the absorbing boundary)(b) $P(\omega)$ for $10^4$ independent RWs, for each case.}
		\label{square_omega}
	\end{figure}

	First, we consider a square lattice ($N = 10^4$ sites) bounded by reflective and absorbing sites, as shown schematically in Fig.~\ref{square_omega} (a). We perform $10^4$ independent RWs, all departing from a common given site $S_0$. In such walks, a time step $t \rightarrow t+1$ consists of a jump from site $S_t$ to a randomly chosen nearest-neighbor site. When a walker tries to jump into a reflective site, it rebounds such that $S_{t+1} = S_t$. On the other hand, when a walker reaches an absorbing site, the walk terminates and the number $\tau$ of steps executed by the walker so far is recorded.
	
	We consider three different starting points on the truss: $ F $ (far from the absorption limit), $ C $ (central lattice) and $ N $ (near the absorption limit). We record the result for $\tau$ and the result for $P(\omega)$ as shown in Fig. 4 (b).

	\begin{figure}[!htp]
	
		\subfloat[]{\includegraphics[width=0.85\linewidth]{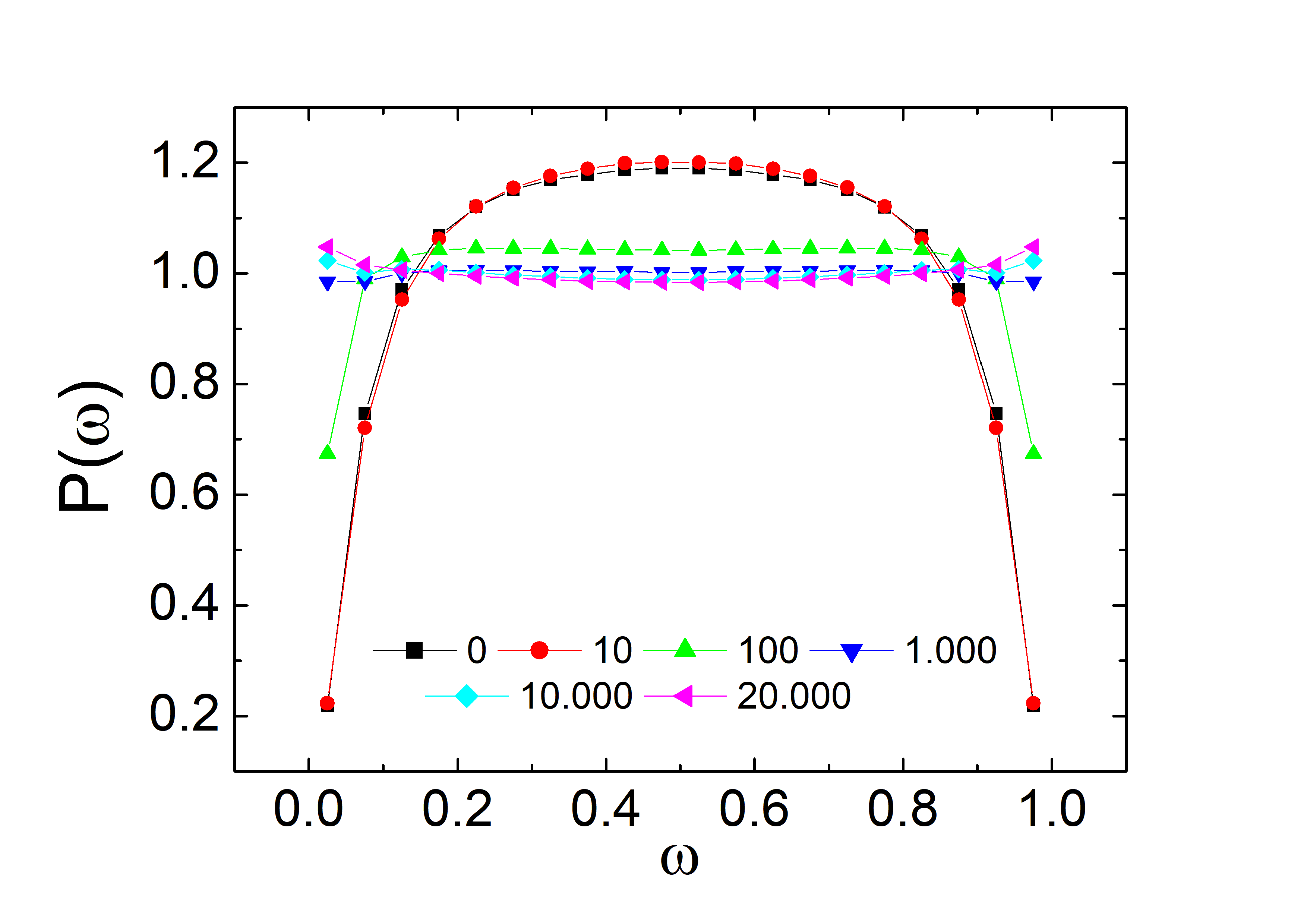}}\\
		\subfloat[]{\includegraphics[width=0.85\linewidth]{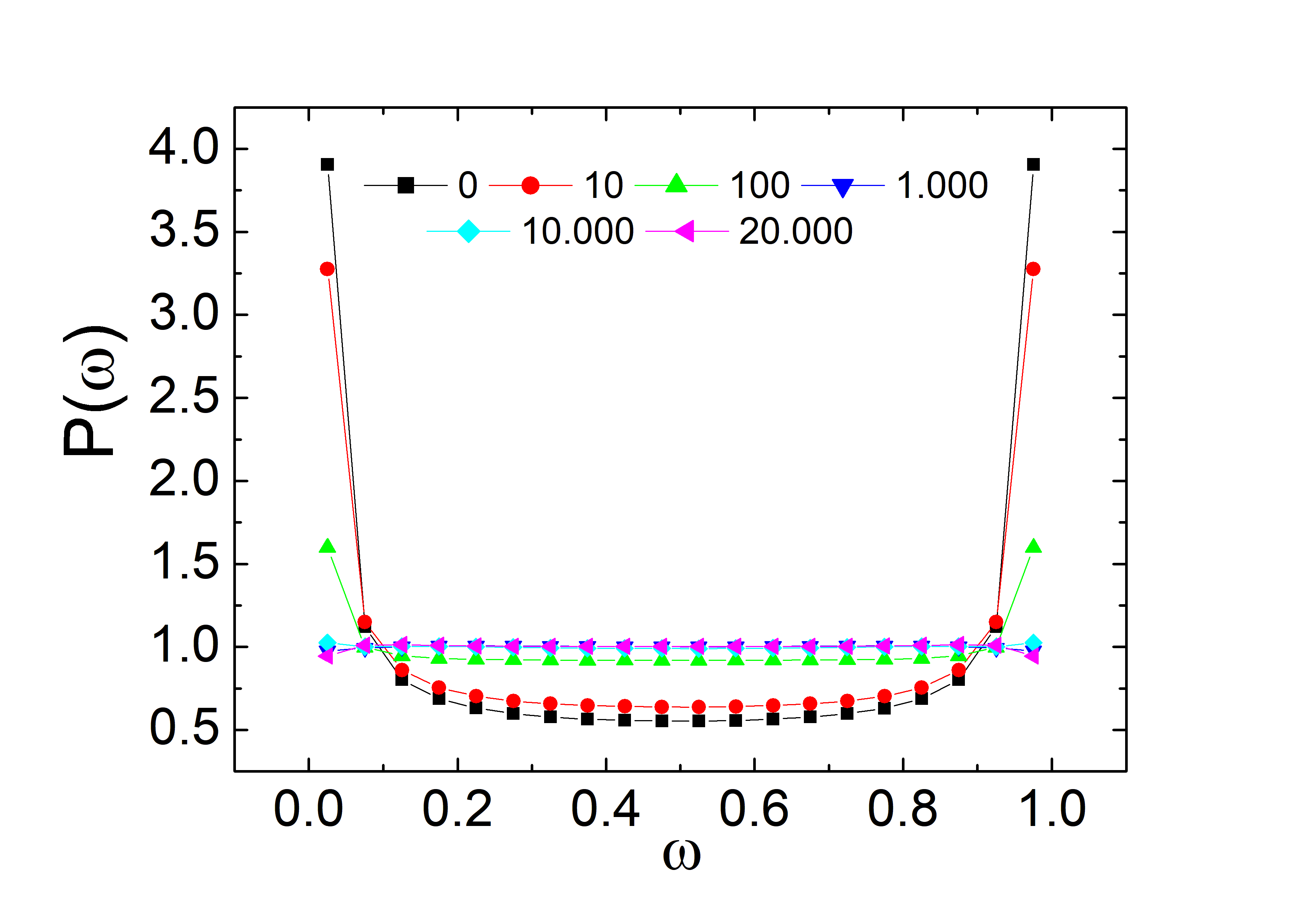}}			 	
		\caption{(Color online) Results in the conservative random network with reflective BCs. $P(\omega)$ for $10^4$ independent RWs, for each case. (a) The target site far from the start site on the network without rewire. (b) The target site near from the start site on the network without rewire. In both cases we increase the number of rewiring between the links.}
		\label{conservative_omega}
	\end{figure}
	
	The results in Fig. \ref{square_omega} (b) show that the change in $P(\omega)$ format depends on the initial location of the random walker in the network. When the walker departs from a location far from an absorbing boundary, the dynamics is uniform. Otherwise, when they depart from a location near an absorbing boundary, the dynamics is very heterogeneous.

\begin{figure}[!htbp] %h or !htbp
		%\centering
		\begin{tabular}{cc}
			\subfloat[0 rewiring]{\includegraphics[width=0.5\linewidth]{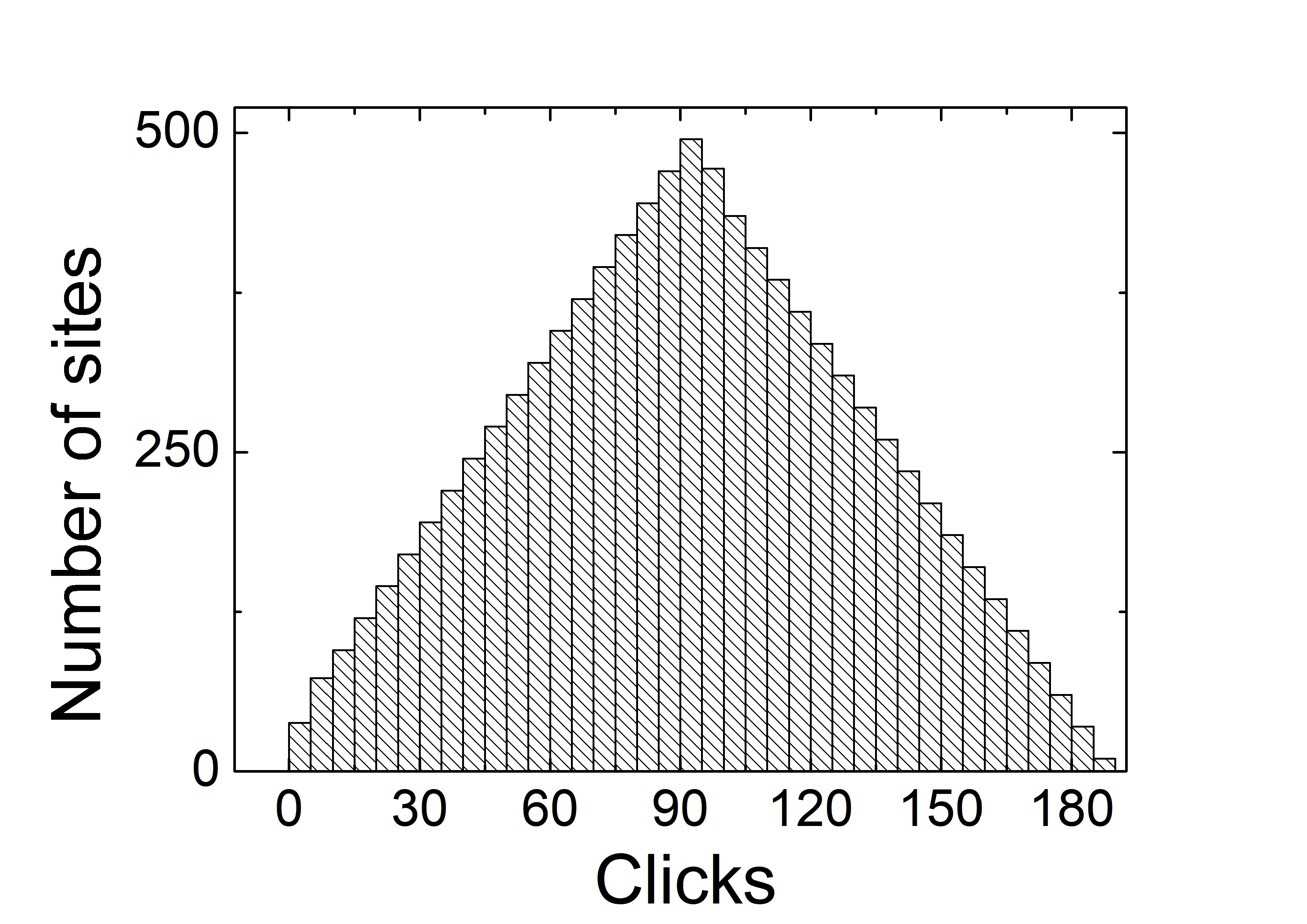}}	 &
			\subfloat[100 rewiring]{\includegraphics[width=0.5\linewidth]{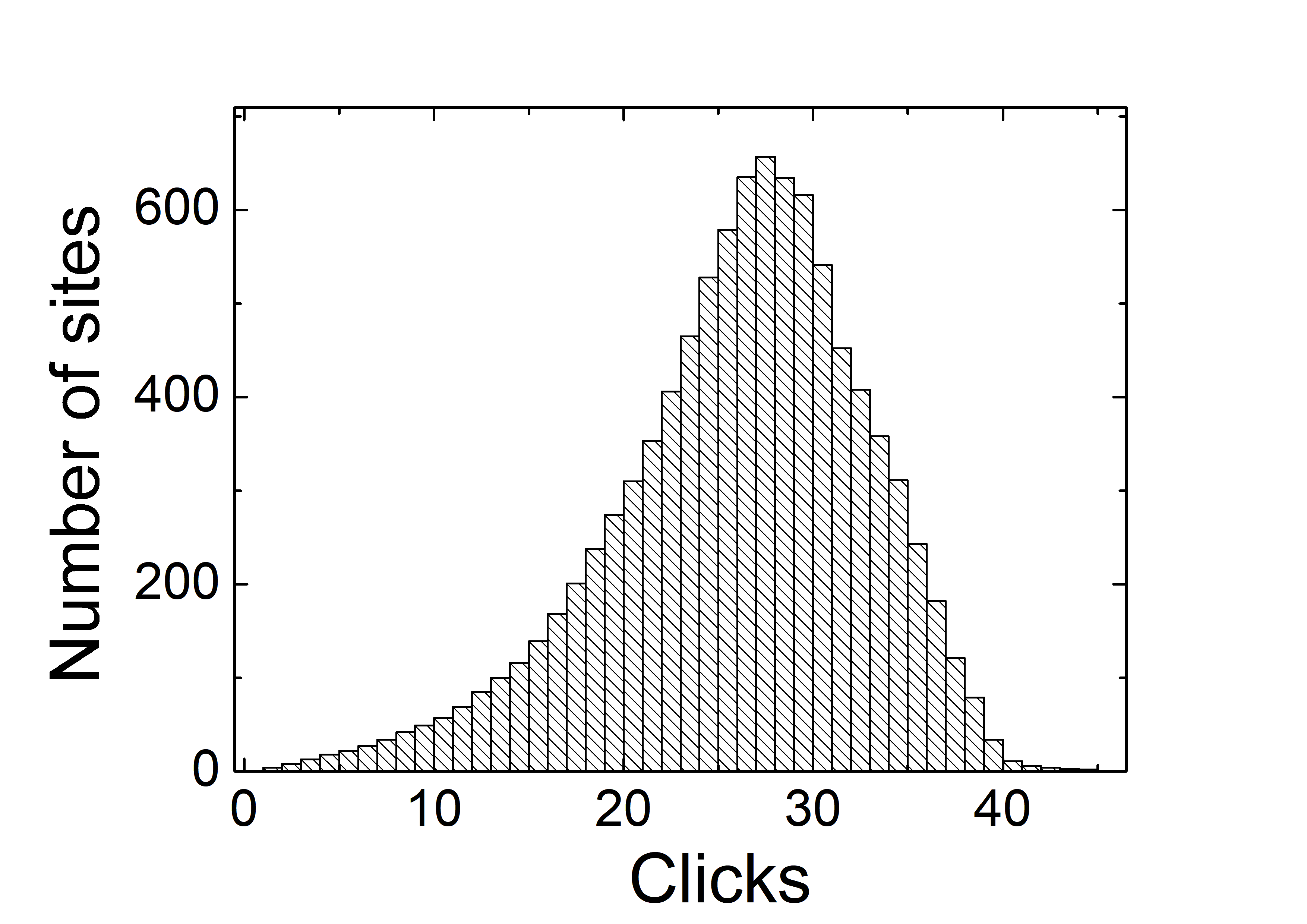}}   \\
			\subfloat[1000 rewiring]{\includegraphics[width=0.5\linewidth]{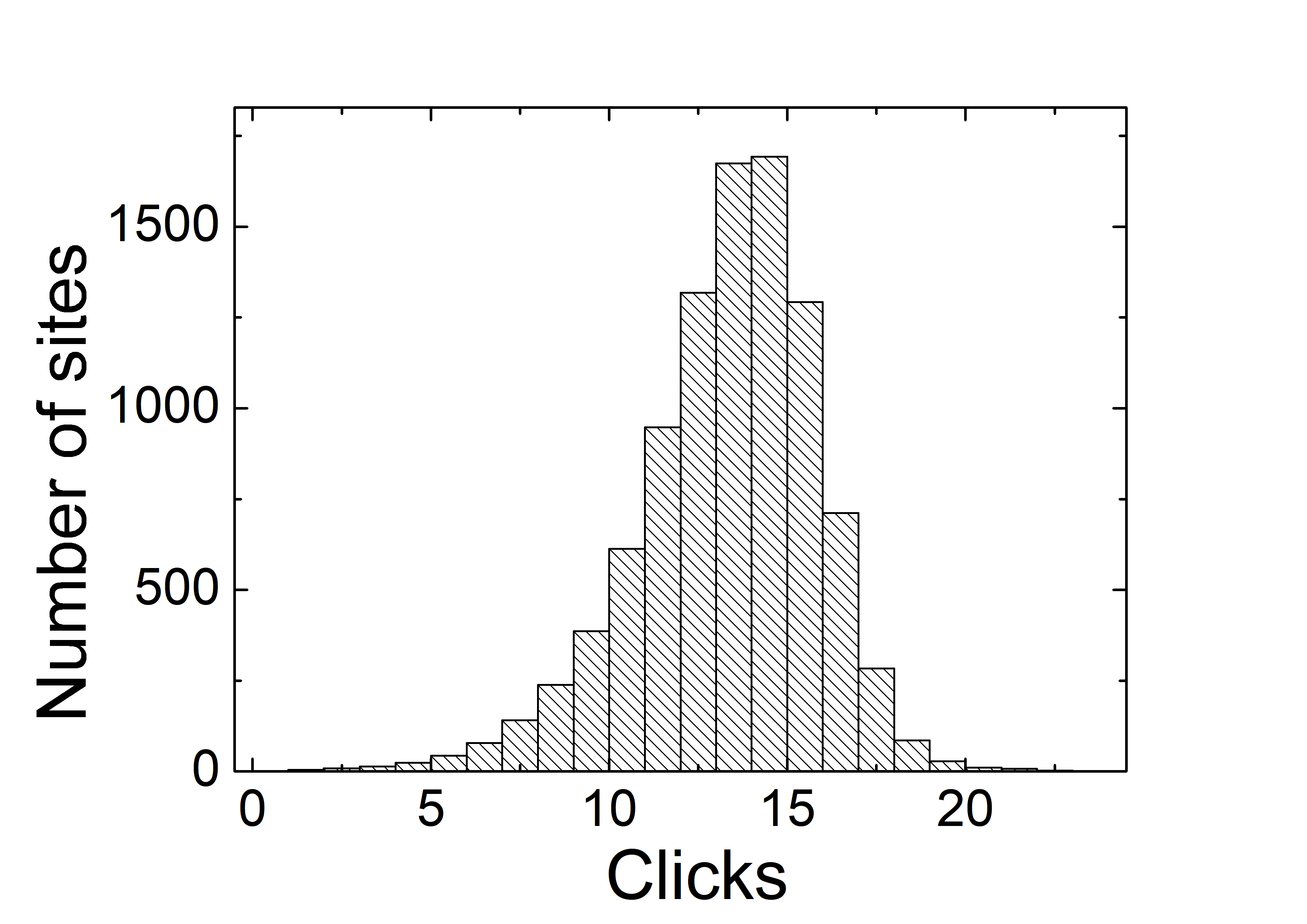}} &
			\subfloat[20000 rewiring ]{\includegraphics[width=0.5\linewidth]{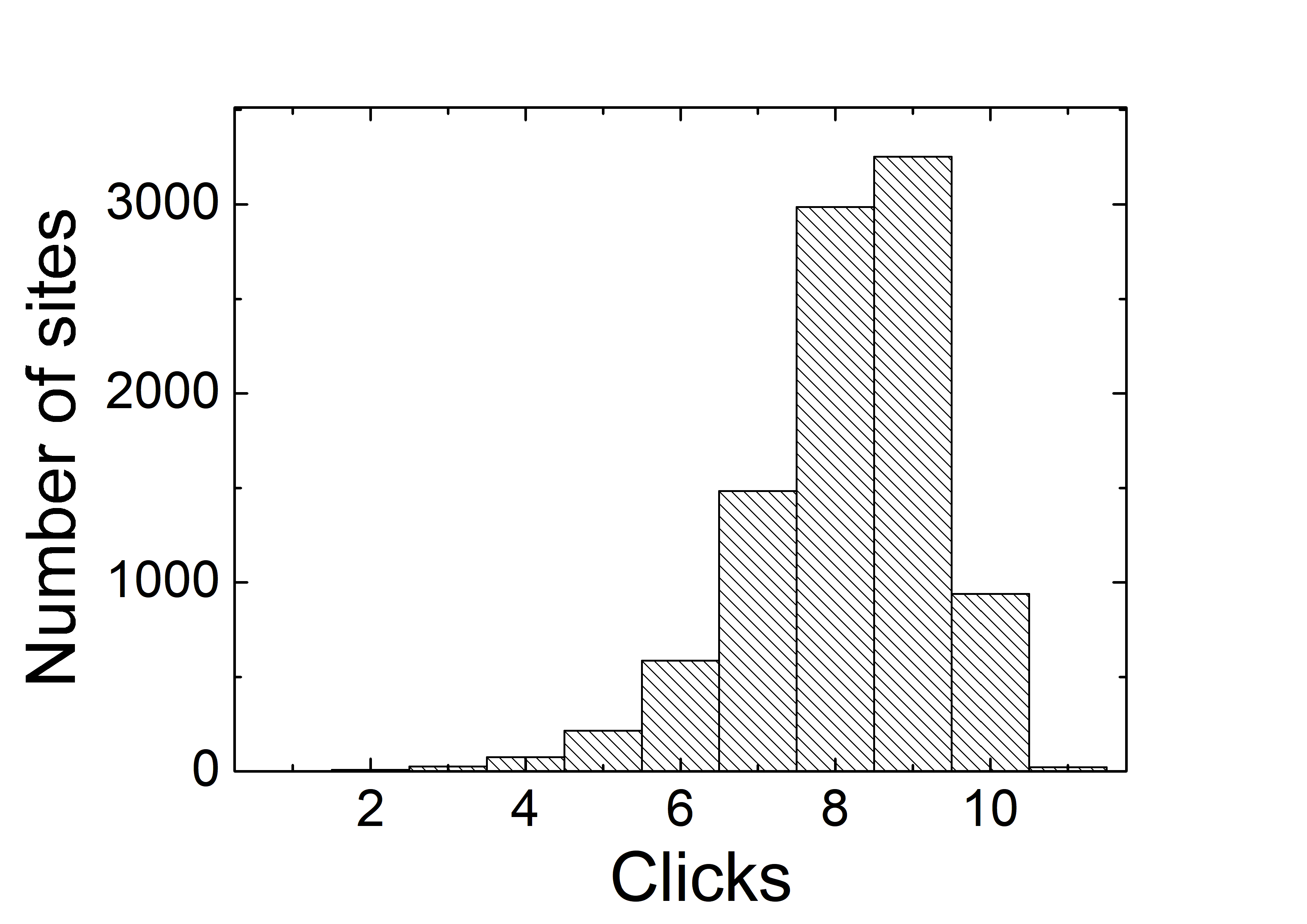}}
		\end{tabular}	
		\caption{Click distribution during the topological change from square network to conservative random network topology, respectively, for 0, 100, 1000 and 20000 rewire of links.}
		\label{crn_clicks}
	\end{figure}

We then consider all sites on the square network boundary as reflective and rewire the links of the initial network to get the conservative random network. The number of clicks separating any site from the target site can be seen in Fig. \ref{crn_clicks}, note the reduction in clicks as links are reconnected. With all links reconnected, the network displays SW features, as shown in Fig. \ref{crn_clicks} (d).

By characterizing the influence of the starting position, $P(\omega)$ it is possible to investigate some topological features, as shown in Fig. \ref{conservative_omega}. While the network is being rewired, the number of clicks between two sites decreases, Fig. \ref{crn_clicks} and $P(\omega)$ change shape, Fig. \ref{conservative_omega}. Note that from $1000$ reconnecting, $ P(\omega)$ becomes stable, indicating that the network topology has changed to SW.

We repeat the analysis on the non-conservative random network. In this case, just over $ 90 \% $ of links on a square network have been reconnected. The results obtained are very similar to those of the conservative random network. Again, starting at $1000$ reconnections, $P(\omega)$ is stable, indicating that the network has begun to exhibit the characteristics of SW.

Next, we investigate the characteristics of the first passage in the SW random network proposed by Watts-Strogatz \cite{Watts_1998}. In this model, we vary the place of departure and fixed the place for arrival.

\begin{figure}[!htp]
	\begin{tabular}{c}
		\subfloat[0 rewiring]{\includegraphics[width=0.85\linewidth]{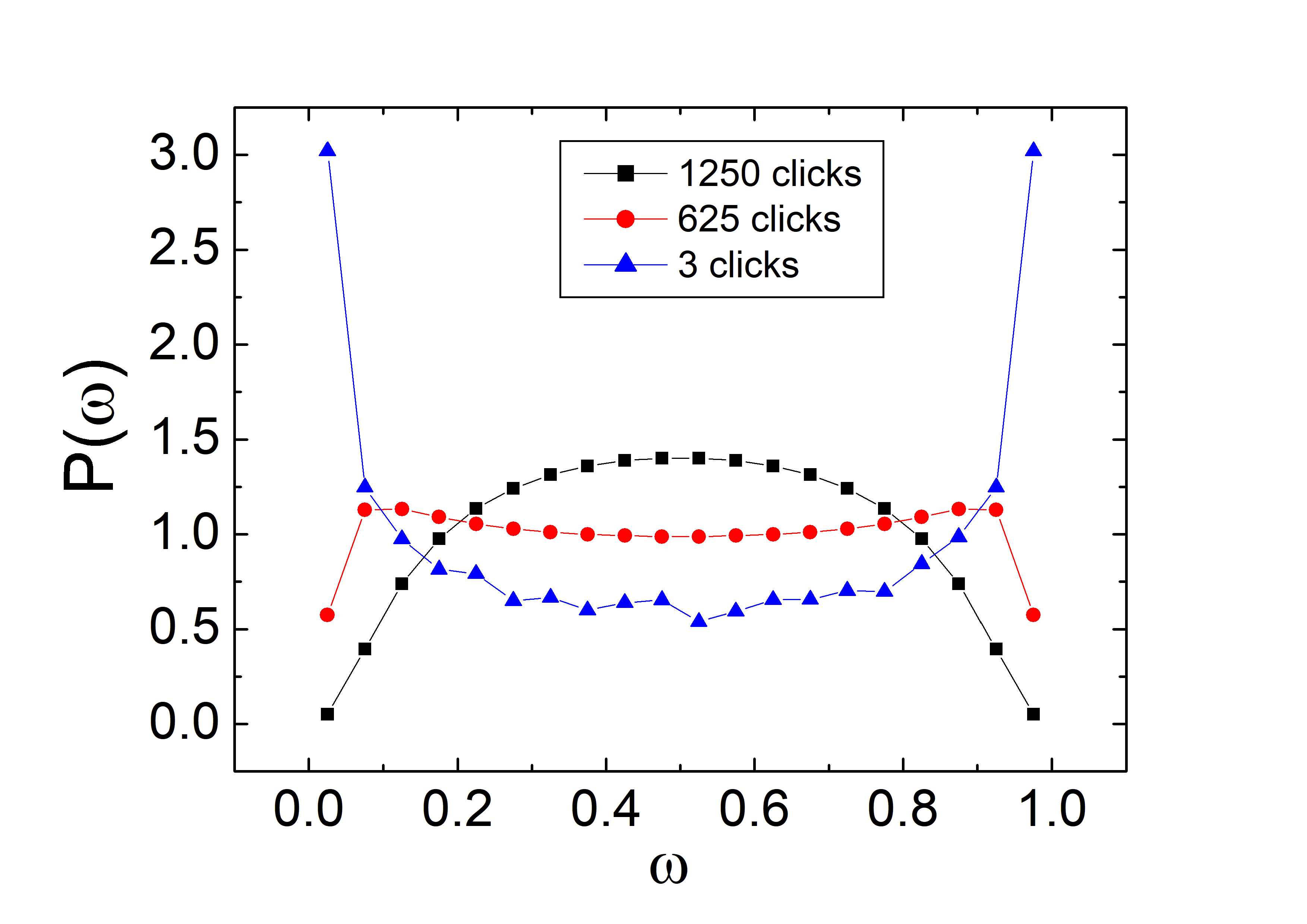}} \\
		\subfloat[20000 rewiring]{\includegraphics[width=0.85\linewidth]{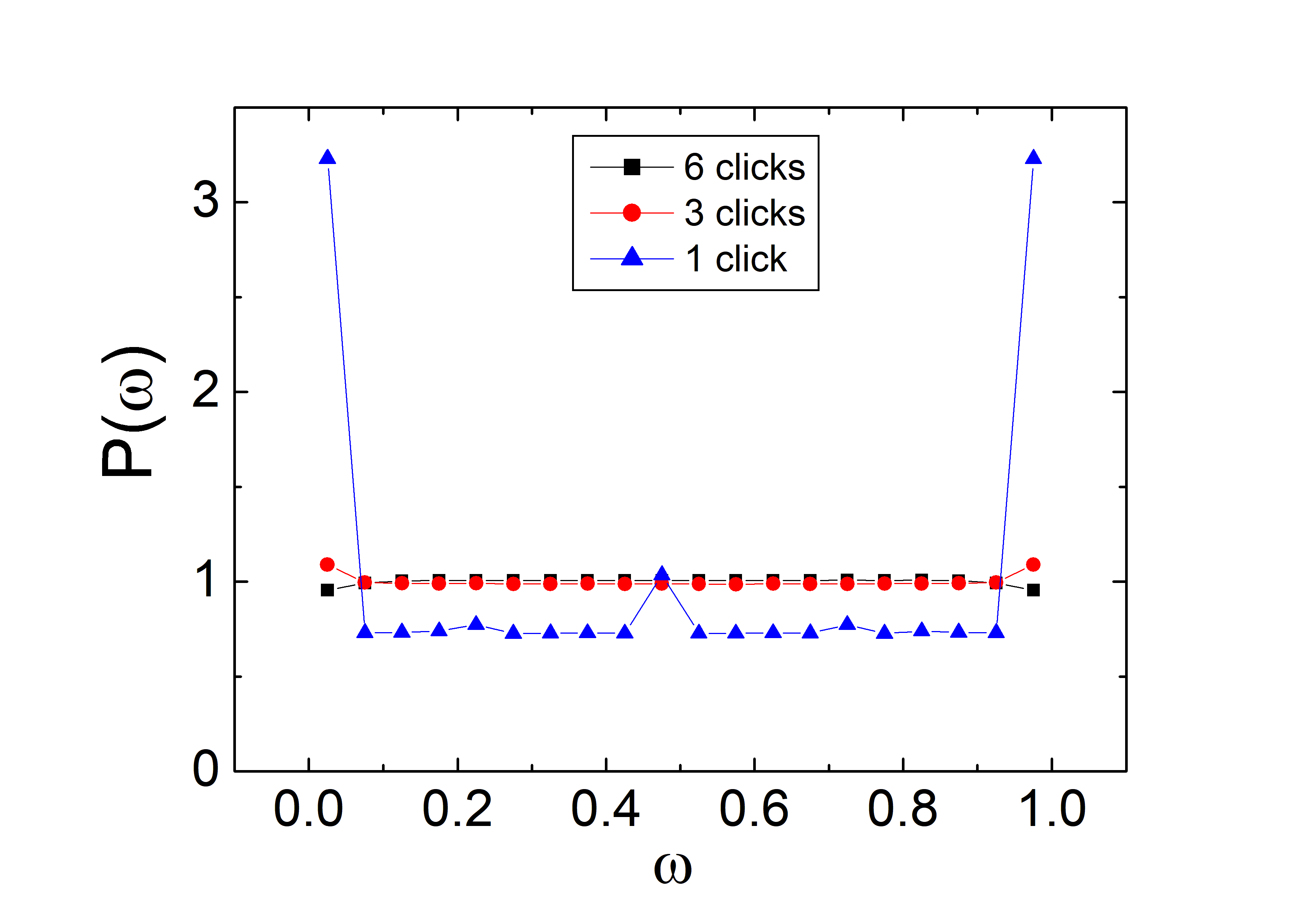}}			 		
	\end{tabular}
	\caption{(Color online) Results in the SW random network. $P(\omega)$ for $10^4$ independent RWs, for each case.}
	\label{ws_omega}
	\end{figure}

First, we investigate a regular network with eight links per site, which can be represented as a ring, without any rewiring of links. Due to network regularity, the three forms of $P(\omega)$ are possible to be identified depending on the initial site, Fig. \ref{ws_omega} (a).

Secondly, we got $P(\omega)$ after the initial ring was completely rewired, Fig. \ref{ws_omega} (b). In this case, the maximum number of clicks between two sites is six and the distribution is essentially flat, both results indicating that the network is a SW.

\begin{figure}[!htp]
		\subfloat[0 rewiring]{\includegraphics[width=0.85\linewidth]{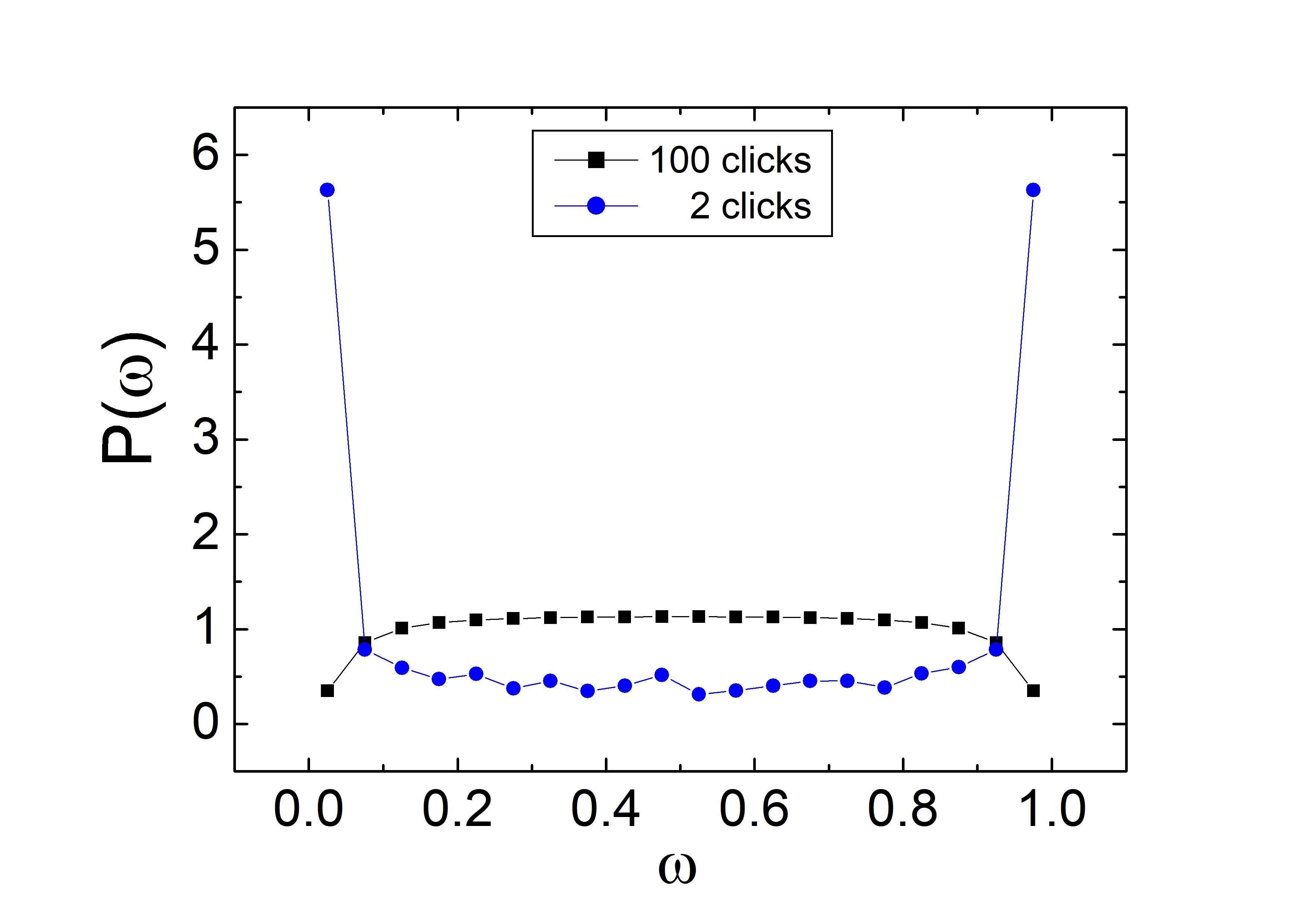}}\\
		\subfloat[18577 rewiring]{\includegraphics[width=0.85\linewidth]{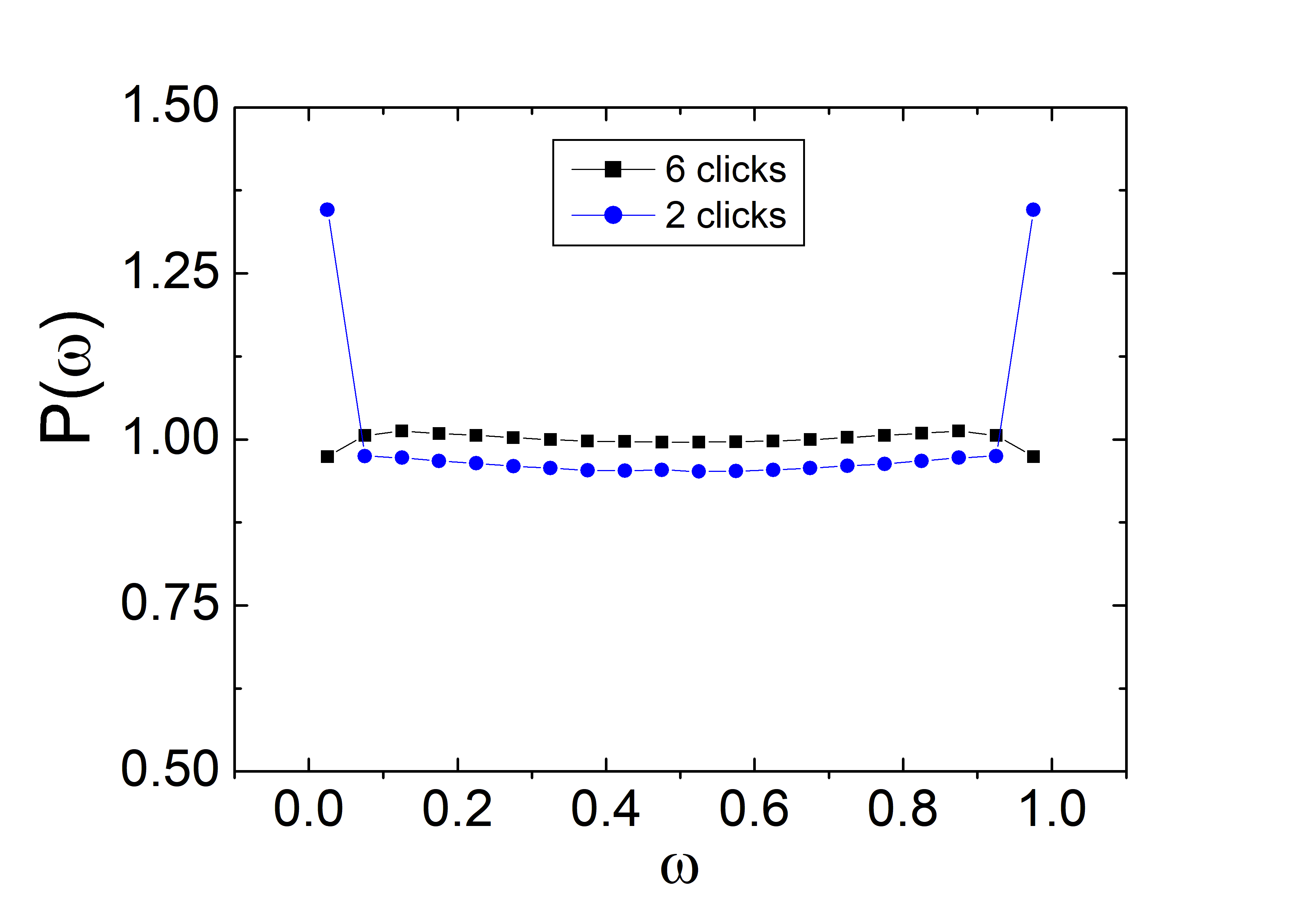}}
		\caption{(Color online) Results in the scale-free network. $P(\omega)$ for $10^4$ independent RWs, for each case.}
	\label{sf_omega}
	\end{figure}

To investigate the scale-free network, we start from the square network without boundary conditions and rewire the links, as described in (\ref{subsec_networks}). Fig. \ref{sf_omega} (a) shows the results for the square network without rewiring its links. In this case, the maximum distance between two sites is 100 clicks. After rebinding the links, the maximum number of clicks between two sites is 6 clicks, and Fig. \ref{sf_omega} (a) displays the behavior in this case. Clearly, we again notice a change in the network topology.
	
$P(\omega)$ has allowed us to identify if a network has the characteristics of SW, but is not able to differentiate between a random network that displays SW properties from a scale-free network.

\begin{figure}[!h] %h or !htbp
	    \centering
   	    \subfloat[SW random network]{\includegraphics[width=0.52\linewidth]{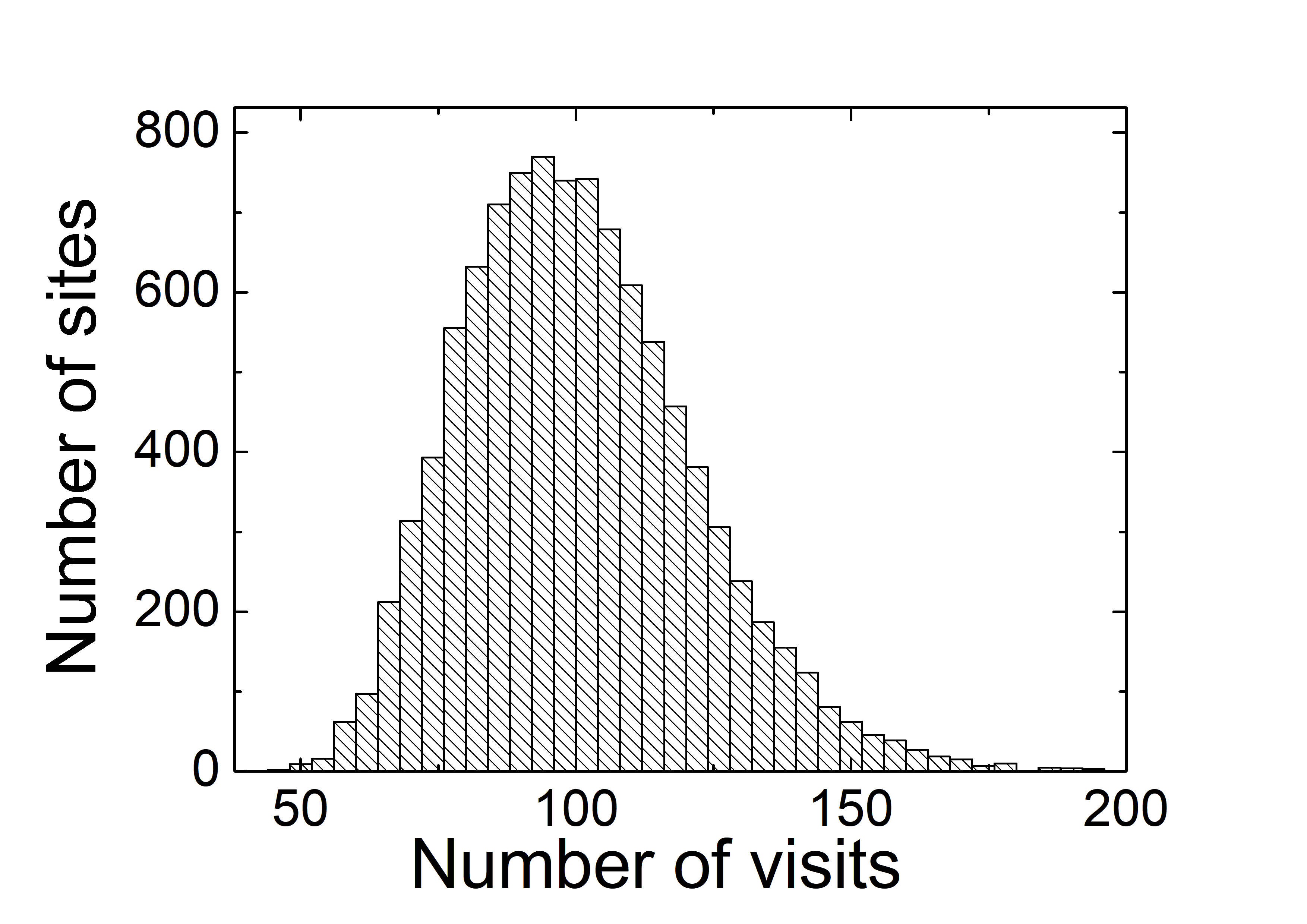}}
		\subfloat[Scale-free network]{\includegraphics[width=0.52\linewidth]{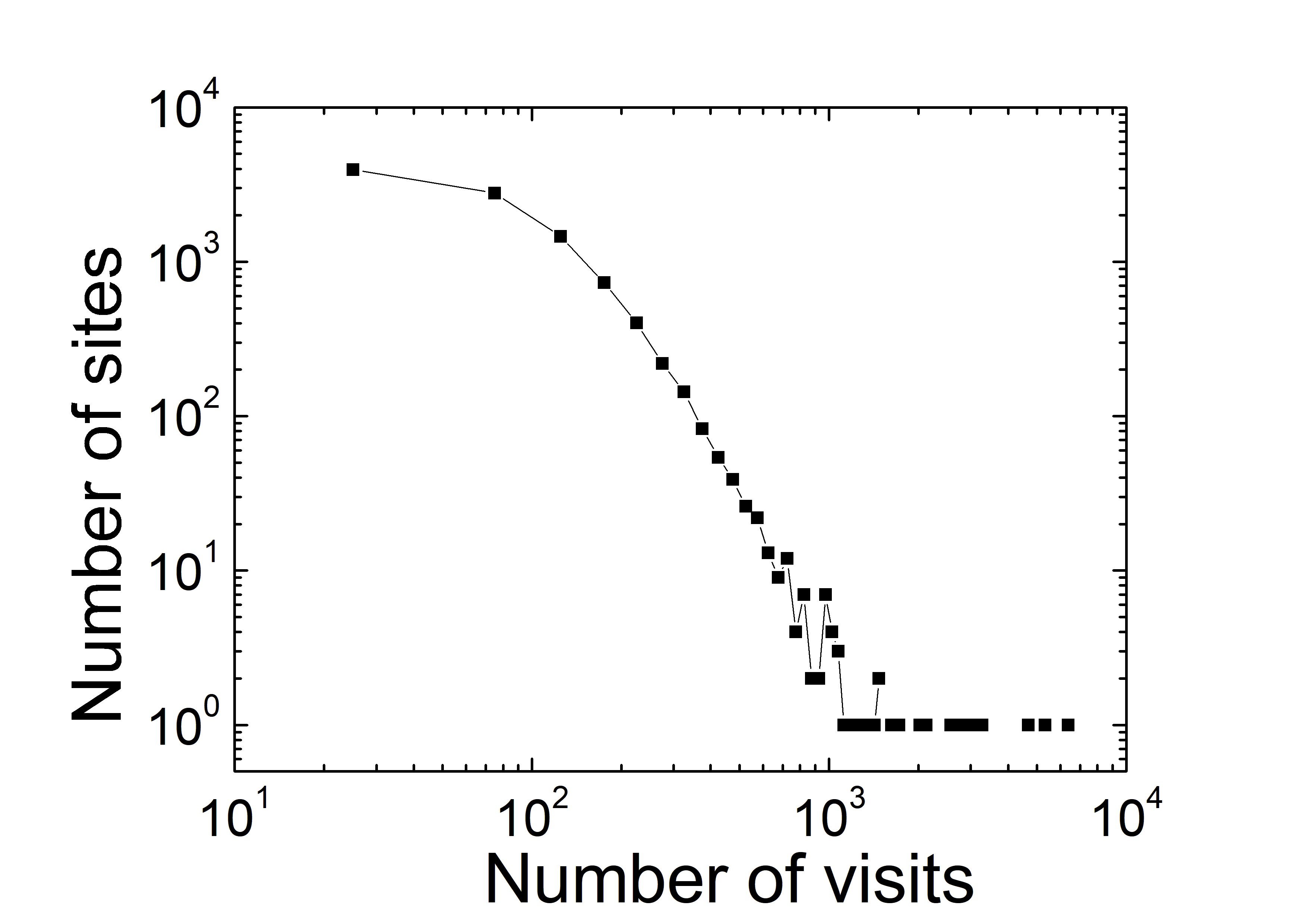}}
		\caption{(Color online) Distribution of occupation of the sites for a random walk with $10^6$ steps, for each case.}
		\label{visits}
	\end{figure}

In order to distinguish such topologies, we analyze the distribution of site occupancy for a random walk with $ 10 ^ 6 $ steps. In the case of the random network, the distribution shows a characteristic number of visits per site, while for the scale-free network, the distribution follows an power law, shown in Fig. \ref {visits}.

%\clearpage
	
\section{Conclusions}
\label{sec:concl}
We have explored the problem of random walks in complex networks. We applied first-passage analysis to characterize different types of networks, such as the conservative and non-conservative random networks, small world networks and scale-free networks. The networks were generated by reconnecting links and different kinds of boundary conditions, such as reflective and absorbing, were considered. By recording first-passage times to an absorbing boundary, we were able to apply first-passage analysis in order to map topological changes in the networks studied, after reconstructing their link structure.

We have shown that first-passage analysis is an efficient tool for characterizing network navigation and topology. We have shown that a change in the shape of the distribution of uniformity index $P(\omega)$ reveals a change in the topology of the network.

The authors wish to thank brazilian funding agencies CAPES and FAPEMIG. APFA thans CNPq grant 308792/2018-1.

%	The clicks on a  lattice are the number of steps to go out a site and arrive another one.  The Fig. \ref{clicks} shows the drastic reduction of the number of clicks required for this, as the links are rewiring using the preferential attachment.

% \section*{References}
\clearpage
\bibliographystyle{apsrev4-1}
\bibliography{./pre}{}

\begin{thebibliography}{99}
%\bibitem{Burkitt_2006} A. N. Burkitt, Biol. Cybern. {\bf 95}, 1 (2006).

\bibitem{Milgram1967}{S. Milgram, Psychology Today 2, 61 (1967).}
\bibitem{Mitchell2009}{M. Mitchell, Complexity (Oxford University Press, 2009).}
\bibitem{Newman2018}{M. Newman, Networks (Oxford University Press, 2018).}
\bibitem{Goh_2001}{K.-I. Goh, B. Kahng, and D. Kim, Physical Review Letters 87 (2001), 10.1103/physrevlett.87.278701.}
\bibitem{Hwang_2013}{S. Hwang, D.-S. Lee, and B. Kahng, Physical Review E 87 (2013), 10.1103/physreve.87.022816.}
\bibitem{Watts_1998}{D. J. Watts and S. H. Strogatz, Nature 393, 440 (1998).}
\bibitem{Hwang_2014}{S. Hwang, D.-S. Lee, and B. Kahng, in First-Passage Phenomena and Their Applications (World Scientific Publishing Company, 2014) pp. 122-144.}
\bibitem{Erdos1959}{P. Erd\"{O}s and A. R\'{E}nyi, Publ. Math. Debrecen 6, 290 (1959).}
\bibitem{Erdos1960}{P. Erd\"{O}s and A. R\'{E}nyi, Publ. Math. Inst. Hungar. Acad. Sci 5, 17 (1960).}
\bibitem{McGraw_2008}{P. N. McGraw and M. Menzinger, Physical Review E 77 (2008), 10.1103/physreve.77.031102.}
\bibitem{Barabasi1999}{A.-L. Barab\'{a}si and R. Albert, Science 286, 509 (1999).}
\bibitem{BARABASI2002}{A.-L. Barab\'{a}si, Linked (Plume Books, 2002).}
\bibitem{Gilbert_1959}{E. N. Gilbert, The Annals of Mathematical Statistics 30, 1141 (1959).}
\bibitem{Goncalves2019}{B. A. Gon\c{c}alves, L. Carpi, O. A. Rosso, M. G. Ravetti, and A. Atman, Physica A: Statistical Mechanics and its Applications 525, 606 (2019).}
\bibitem{Carpi2019}{L. C. Carpi, T. A. Schieber, P. M. Pardalos, G. Marfany, C. Masoller, A. D\'{\i}az-Guilera, and M. G. Ravetti, Scientific Reports 9 (2019), 10.1038/s41598-019-38869-0.}
\bibitem{Oliveira_2019}{I. M. de Oliveira, L. Carpi, and A. P. F. Atman, Under review (2019).}
\bibitem{FPP2014}{R. Metzler, G. Oshanin, and S. Redner, eds., First-passage phenomena and their applications (World Scientific Publishing Company, 2014).}
\bibitem{Chou2014}{T. Chou and M. R. D'Orsogna, in First-Passage Phenomena and Their Applications (World Scientific Publishing Company, 2014) Chap. 13, pp. 306-345.}
\bibitem{Choe2019}{G. H. Choe, H. J. Jang, and Y. H. Na, Statistics \& Probability Letters 148, 43 (2019).}
\bibitem{Burkitt_2006}{A. N. Burkitt, Biological Cybernetics 95, 1 (2006).}
\bibitem{Tejedor_2009}{V. Tejedor, O. B\'{e}nichou, and R. Voituriez, Physical Review E 80 (2009), 10.1103/physreve.80.065104.}
\bibitem{Tejedor_2010}{V. Tejedor, O. B\'{e}nichou, R. Voituriez, and M. Moreau, Physical Review E 82 (2010), 10.1103/physreve.82.056106.}
\bibitem{Rosvall_2008}{M. Rosvall and C. T. Bergstrom, Proceedings of the National Academy of Sciences 105, 1118 (2008).}
\bibitem{Montroll1964}{E. W. Montroll, Proceedings of Symposia in Applied Mathematics 16, 193 (1964).}
\bibitem{Montroll_1965}{E. W. Montroll and G. H. Weiss, Journal of Mathematical Physics 6, 167 (1965).}
\bibitem{Condamin2005}{S. Condamin, O. B\'{e}nichou, and M. Moreau, Physical Review Letters 95 (2005), 10.1103/physrevlett.95.260601.}
\bibitem{Benichou2010}{O. B\'{e}nichou, C. Chevalier, J. Klafter, B. Meyer, and R. Voituriez, Nature Chemistry 2, 472 (2010).}
\bibitem{Noh_2004}{J. D. Noh and H. Rieger, Physical Review Letters 92 (2004), 10.1103/PhysRevLett.92.118701.}
\bibitem{Margolus1987}{N. M. Tommaso Toffoli, Cellular Automata Machines (MIT Press Ltd, 1987).}
\bibitem{Mejia-Monasterio2011}{C. Mej\'{\i}a-Monasterio, G. Oshanin, and G. Schehr, Journal of Statistical Mechanics: Theory and Experiment 2011, P06022 (2011).}
\bibitem{Havlin_2002}{S. Havlin and D. Ben-Avraham, Advances in Physics 51, 187 (2002).}
\bibitem{Redner2007}{S. Redner, A Guide to First-Passage Processes (Cambridge University Press, 2007).}
\bibitem{Reis2014}{F. D. A. A. ao Reis and D. di Caprio, Physical Review E 89 (2014), 10.1103/PhysRevE.89.062126.}
\bibitem{Metzler_2000}{R. Metzler and J. Klafter, Physica A 278, 107 (2000).}
\bibitem{Scher_2002}{H. Scher, G. Margolin, R. Metzler, J. Klafter, and B. Berkowitz, Geophys. Res. Lett. 29, 1061 (2002).}
\bibitem{Mattos_2012}{T. G. Mattos, C. Mej\'{\i}a-Monasterio, R. Metzler, and G. Oshanin, Physical Review E 86 (2012), 10.1103/PhysRevE.86.031143.}
\bibitem{Mattos_2014}{T. G. Mattos, C. Mej\'{\i}a-Monasterio, R. Metzler, G. Oshanin, and G. Schehr, in First-Passage Phenomena and Their Applications (World Scientific Publishing Company, 2014) Chap. 9, pp. 203-225.}

\end{thebibliography}

\end{document}